\documentclass[prl,twocolumn,aps,longbibliography,superscriptaddress,notoc]{revtex4-1}
\usepackage{amsmath}
\usepackage{amssymb}
\usepackage{bm}
\usepackage{epsfig}
\usepackage{graphicx}
\usepackage[dvipsnames]{xcolor}
\usepackage{color}
\usepackage[unicode=true,colorlinks=true,citecolor=blue,urlcolor=blue]{hyperref}
\usepackage[english]{babel}
\usepackage{braket}

\let\ifr\i

\renewcommand{\Re}{\mathop{\rm Re}}

\renewcommand{\sinh}{\mathop{\rm sh}}
\renewcommand{\cosh}{\mathop{\rm ch}}

\newcommand{\e}{\mathrm{e}}

\renewcommand{\i}{{\rm i}}
\renewcommand{\d}{\mathrm d}
\renewcommand{\emph}{\textit}

\renewcommand{\braket}[1]{\left\langle #1 \right\rangle}

\newcommand{\addDima}[1]{#1}

\newcommand{\nix}[1]{}
\let\oldsec\section
\renewcommand{\section}[1]{\textit{#1}---}

\begin{document}

\title{Nuclear spin dynamics, noise, squeezing and entanglement in box model}

\author{A.~V.~Shumilin}
\author{D.~S.~Smirnov}
\email{smirnov@mail.ioffe.ru}
\affiliation{Ioffe Institute, 194021 St. Petersburg, Russia}

\date{\today}

\begin{abstract}
We obtain a compact analytical solution for the nonlinear equation for the nuclear spin dynamics in the central spin box model in the limit of many nuclear spins. The total nuclear spin component along the external magnetic field is conserved and the two perpendicular components precess or oscillate depending on the electron spin polarization, with the frequency, determined by the nuclear spin polarization. \addDima{As applications of our solution, we calculate the nuclear spin noise spectrum and describe the effects of nuclear spin squeezing and many body entanglement in the absence of a system excitation.}
\end{abstract}

\maketitle

\section{Introduction}The problem of a single ``central'' spin interaction with surrounding spins is known as the central spin model. It is widely used to describe the interaction of a localized electron with nuclei, for example, in quantum dots or in the vicinity of donors in bulk semiconductors~\cite{book_Glazov}. Generally, this is a complex many body problem, and it was studied in many details~\cite{Coisch_review,Yang_2016}. In particular, the central spin model allows one to describe the 
Hanle effect in transverse magnetic field~\cite{PhysRevB.67.073301}, polarization recovery in longitudinal field~\cite{PhysRevLett.94.116601,PRC}, spin precession mode locking~\cite{yugova12}, nuclei-induced frequency focusing~\cite{A.Greilich09282007}, spin noise~\cite{NoiseGlazov,PhysRevLett.109.166605,SmirnovUFN}, effect of the spin inertia~\cite{Korenev2015,PhysRevB.98.125306}, dynamic nuclear spin polarization~\cite{OptOr} and many other effects.

\addDima{The interest in the interwinded electron and nuclear spin dynamics is mostly driven by the perspective of quantum dots based scalable technology for quantum computations~\cite{PhysRevLett.91.246802,Reilly817,Chekhovich_protocol}. Most of the previous studies considered the nuclear spins as an important source of electron spin decoherence~\cite{merkulov02,PhysRevLett.88.186802,PhysRevB.70.195340}. But recently the nuclear spins were recognized as a possible platform for quantum information storage and procession~\cite{PhysRevLett.90.206803,discretization}. For example, coherent interface between electron and nuclear spins was developed~\cite{Gangloff62}, sensing of single quantum nuclear spin excitation was realized~\cite{Jackson_sensing}, elementary quantum algorithms were implemented in the nuclear spin quantum register in strained quantum dots~\cite{Chekhovich_register}.}




The complexity of the nuclear spin dynamics is related mainly with a huge number of nuclei interacting with a single electron. Despite the possibility to diagonalize the central spin model Hamiltonian for a finite number of nuclei using the Bethe ansatz~\cite{Gaudin} and to calculate the nuclear spin dynamics in the box model~\cite{PhysRevB.70.014435,Kozlov2007,Bortz_2007}, it is still hardly possible to qualitatively describe the nuclear spin dynamics especially for many nuclear spins~\cite{Chertkov,PhysRevB.51.3974,PhysRevLett.124.196801}. In this work we solve this long standing problem and obtain the exact expressions for the nuclear spins dynamics in the limit of many nuclear spins. \addDima{These expressions are used to calculate the nuclear spin noise spectra, and to describe the effects of intrinsic nuclear spin squeezing and many body entanglement in the central spin model.}

\section{Nuclear spin dynamics in the box model}The Hamiltonian of the box model has the form
\begin{equation}
  \label{eq:Ham}
  \mathcal H=A\bm I\bm S+\hbar\bm\Omega_B\bm S+\hbar\bm\omega_B\bm I,
\end{equation}
where $A$ is the constant of the hyperfine coupling between the total nuclear spin $\bm I$ and the electron spin $\bm S$, and $\bm\Omega_B$ and $\bm\omega_B$ are electron and nuclear spin precession frequencies in the external magnetic field, respectively. Throughout the paper we use the minuscule and majuscule omegas to denote the nuclear and electron spin precession frequencies, respectively. The total nuclear spin is composed of $N$ of individual nuclear spins $\bm I_n$:
$
  \bm I=\sum_{n=1}^N\bm I_n.
$
Thus the box model is a particular case of the central spin model, where all the hyperfine coupling constants are equal.

\begin{figure}[t]
  \centering
  \includegraphics[width=\linewidth]{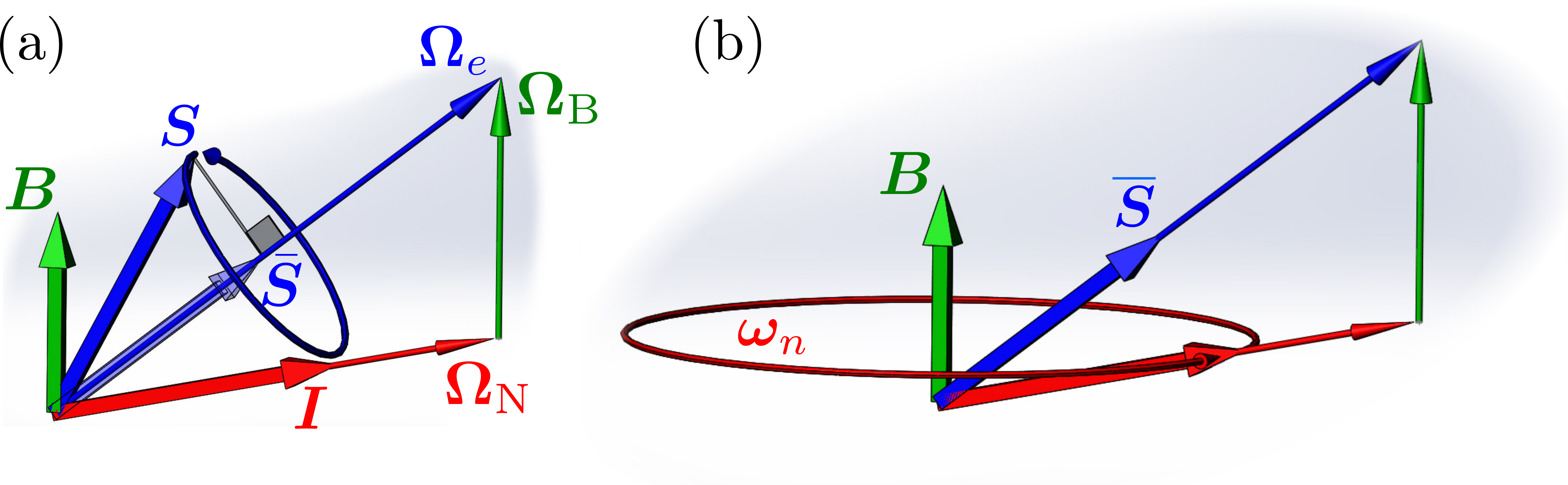}
  \caption{(a) Electron spin precesses around the sum of the external magnetic field and the Overhauser field, and effectively projects to the direction of $\bm\Omega_e$. (b) At the long time scales the average electron spin adiabatically follows the direction of $\bm\Omega_e$ and induces the nuclear spin precession around the direction of the magnetic field with the frequency $\bm\omega_n$.}
  \label{fig:precession_both}
\end{figure}

In the Heisenberg representation the electron spin operator obeys the Bloch equation
\begin{equation}
  \label{eq:spin_prec}
  \frac{\d\bm S}{\d t}=\bm\Omega_e\times\bm S,
\end{equation}
where
$
  \bm\Omega_e=\bm\Omega_B+\bm\Omega_N,
$
is the total electron spin precession frequency with
$
  \bm\Omega_N=A\bm I/\hbar
$
being the frequency related to the Overhauser field. Thus the electron spin rotates in the sum of the nuclear and external magnetic fields, as illustrated in Fig.~\ref{fig:precession_both}(a).

Similarly, the nuclear spin operator obeys
\begin{equation}
  \label{eq:nucl_prec}
  \frac{\d\bm I}{\d t}=\left(\frac{A}{\hbar}\bm S+\bm\omega_B\right)\times\bm I.
\end{equation}
One can see that, the system states with the different absolute values of the total nuclear spin $I$ are not mixed, so it is a good quantum number.

\addDima{Generally, one can not replace in Eqs.~\eqref{eq:spin_prec} and~\eqref{eq:nucl_prec} the operators $\bm S$ and $\bm I$ with their average values and solve the resulting equations. This procedure would give, for example, no effect of the hyperfine interaction for the nuclear spin dynamics if the electron spin is unpolarized, which is not correct because of the quantum uncertainty for electron spin, which is as large as its maximum possible average value. This problem was solved previously only numerically. Below we solve it analytically in the limit of a large nuclear spin.}


In self assembled GaAs quantum dots, typically $N\sim10^5$, so even in the absence of nuclear spin polarization the typical fluctuation of $I\sim\sqrt{N}$ is very large. Note also that the nuclear magnetic moment is much smaller than that of electron, so we assume that $\omega_B\ll\Omega_B$. In this case the electron spin precession is much faster than that of the nuclei~\cite{merkulov02}, which allows us to find the compact exact solution.

Formally, the solution of Eq.~\eqref{eq:spin_prec} is
$
  \bm S(t)=\e^{\i\mathcal Ht/\hbar}\bm S\e^{-\i\mathcal Ht/\hbar}.
$
For large nuclear spin $I\gg 1$ we neglect the commutator of its components hereafter~\footnote{This can be done at the short time scales, when the nuclear spin dynamics can be neglected. However, the closed set of equations for the nuclear spin dynamics obtained below, is exact and can be used for the time scales of the order of $1/\omega_e$.} (it was not neglected in the derivation of Eq.~\eqref{eq:nucl_prec} for the only time), which yields
$
  \bm S(t)=\e^{\i\bm\Omega_e\bm St}\bm S\e^{-\i\bm\Omega_e\bm S t}.
$
The standard decomposition of the spin matrix exponents gives
\begin{multline}
  \label{eq:S3}
  \bm S(t)=\left[\cos(\Omega_et/2)+2\i\frac{\bm S\bm\Omega_e}{\Omega_e}\sin(\Omega_e t/2)\right]\bm S\\\times\left[\cos(\Omega_et/2)-2\i\frac{\bm S\bm\Omega_e}{\Omega_e}\sin(\Omega_e t/2)\right].
\end{multline}
Note that $\bm\Omega_e$ here is still an operator. In fact this expression contains only the even powers of $\Omega_e$, which can be calculated as $\Omega_e^2=\bm \Omega_e^2$.

Eq.~\eqref{eq:S3} contains oscillating terms and has nonzero time average
\begin{equation}
  \bar{\bm S}=\frac{\bm\Omega_e(\bm\Omega_e\bm S)}{\Omega_e^2}.
\end{equation}
It has the meaning of the projection of the electron spin on the direction of $\bm\Omega_e$~\cite{merkulov02}, as illustrated Fig.~\ref{fig:precession_both}(a). Note that $\bar{\bm S}$ is an operator and not a quantum mechanical average.

In view of the separation of the time scales of the electron and nuclear spin dynamics, the electron spin in Eq.~\eqref{eq:nucl_prec} can be replaced with its average:
\begin{equation}
  \label{eq:dI}
  \frac{\d\bm I}{\d t}=\left(\frac{A}{\hbar}\bar{\bm S}+\bm\omega_B\right)\times\bm I.
\end{equation}
It is convenient to rewrite this equation as
\begin{equation}
  \label{eq:dI2}
  \frac{\d\bm I}{\d t}=\frac{A}{\hbar}\bm e_z\times\bm J +\bm\omega_B\times\bm I,
\end{equation}
where
$
  \bm J =(\bm\Omega_e\bm S)\Omega_B\bm I/\Omega_e^2
$
describes the correlation between electron and nuclear spins and $\bm e_z$ is the unit vector along $\bm\Omega_B$ direction. \addDima{This is an auxiliary quantity. For example, in strong magnetic field $\bm\Omega_e\approx\bm\Omega_B$, so $\bm J=\bm IS_z$, which has a clear meaning of the electron and nuclear spins correlator.}

We note that $\mathcal H\approx\bm\Omega_e\bm S$, so this product is constant, which can be called the adiabatic approximation. Moreover, $\mathcal H^2\approx\Omega_e^2/4$, so $\Omega_e^2$ is also constant. Therefore, using Eq.~\eqref{eq:nucl_prec} we obtain
\begin{equation}
  \label{eq:dI_tilde}
  \frac{\d\bm J }{\d t}=\frac{A}{\hbar}\frac{\Omega_B^2}{4\Omega_e^2}\bm e_z\times\bm I+\bm\omega_B\times\bm J .
\end{equation}
This equation along with Eq.~\eqref{eq:dI2} forms a closed set. It accounts for the electron spin commutation relations, but neglects the nuclear ones. This set is exact in the limit of large $I$, and this is the main result of this work.

\section{Quasiclassical interpretation}In Eqs.~\eqref{eq:dI2} and~\eqref{eq:dI_tilde} all the quantities (except for $\bm\Omega_B$ and $\bm\omega_B$) are operators. In this section we replace all the operators with their average values, but use the same notations for brevity.

It is convenient to rewrite Eqs.~\eqref{eq:dI2} and~\eqref{eq:dI_tilde} for the quantum mechanical average values in more physically transparent notations. The direction of $\bm\Omega_e$ represents a good electron spin quantization axis, so the quantities
$
  P_\pm=1/2\pm\bm\Omega_e\bm S/\Omega_e
$
represent the probabilities for the electron spin to be parallel or antiparallel to this direction. We also introduce
\begin{equation}
  \label{eq:Ipm}
  \bm I^\pm=\left(\frac{\bm I}{2}\pm\frac{\Omega_e}{\Omega_B}\bm J\right)/P_\pm,
\end{equation}
which represent the nuclear spin in these two cases, respectively. Importantly, one should use the average value $\bm J$ here and one should not replace it with the product of the average values from the definition in order to correctly describe the correlations between electron and nuclear spins. The total nuclear spin is given by
$
  \bm I=P_+\bm I^++P_-\bm I^-.
$

From Eqs.~\eqref{eq:dI2} and~\eqref{eq:dI_tilde} we simply obtain
\begin{equation}
  \label{eq:dIpm0}
  \frac{\d\bm I^\pm}{\d t}=\bm\omega_n^\pm\times\bm I^\pm,
\end{equation}
where
\begin{equation}
  \label{eq:omega_n_f}
  \bm\omega_n^\pm=\pm\omega_e\frac{\bm\Omega_B}{\Omega_e}+\bm\omega_B,
\end{equation}
with $\omega_e=A/(2\hbar)$ being the nuclear spin precession frequency in the Knight field of completely spin polarized electron. So in the cases of the electron spin parallel or antiparallel to $\bm\Omega_e$, the total nuclear spin precesses with the frequency $\bm\omega_n^\pm$, respectively. It is illustrated in Fig.~\ref{fig:precession_both}(b). The external magnetic field tilts the average electron spin $\bar{\bm S}$ from the direction of $\bm\Omega_N$ to $\bm\Omega_e$. As a result, the Knight field being parallel to $\bar{\bm S}$ tilts from the direction of $\bm I$ and leads to the nuclear spin precession~\cite{yugova11}. However, this precession is slow, so the electron spin adiabatically follows the direction of $\bm\Omega_e$. In this case the Knight, Overhauser and external magnetic field always lie in the same plane, so the nuclear spin rotates around the $z$ axis with the frequency $\bm\omega_n^\pm$. The total nuclear spin dynamics represents the superposition of precessions with these two frequencies.
 We stress that due to the dependence of $\bm\omega_n^\pm$ on $\bm\Omega_N$, equations~\eqref{eq:dIpm0} describing the nuclear spin dynamics are foramally nonlinear.


The solution of Eqs.~\eqref{eq:dIpm0} in the case of $\omega_B=0$ yields
\begin{subequations}
  \label{eq:nucl_osc}
  \begin{equation}
    I_x(t)=I_x(0)\cos(\omega_n t)-\frac{2(\bm\Omega_e\bm S)}{\Omega_e}I_y(0)\sin(\omega_n t),
  \end{equation}
  \begin{equation}
    I_y(t)=I_y(0)\cos(\omega_n t)+\frac{2(\bm\Omega_e\bm S)}{\Omega_e}I_x(0)\sin(\omega_n t),
  \end{equation}
\end{subequations}
where $\omega_n=|\bm\omega_n^\pm|$ (note that $\bm\Omega_e\bm S$ and $\Omega_e$ do not depend on time). Crucially, these expressions demonstrate that the nuclear spin oscillates even in the absence of the electron spin polarization ($\bar{\bm S}=0$) due to \addDima{the electron spin quantum uncertainty. In this case the superposition of the two precessions in the Knight field with the opposite frequencies result in the nuclear spin oscillations, while $\omega_B=0$.}

We have checked that our theory agrees with the numerical solution of the Schr\"odinger equation with the accuracy $\propto1/N$~\cite{supp}.


\section{Nuclear spin noise}The nuclear spin dynamics can be most easily studied experimentally in close to equilibrium conditions through its action on electron. In this case it is characterized by the nuclear spin noise spectra~\cite{SmirnovUFN,PhysRevB.97.195311}
\begin{equation}
  \label{eq:Sz_def}
  (I_\alpha^2)_\omega=\int\limits_{-\infty}^\infty\braket{I_\alpha(t)I_\alpha(t+\tau)}\e^{\i\omega\tau}\d\tau,
\end{equation}
where the angular brackets denote the statistical averaging. These spectra can be measured directly using the resonance shift spin noise spectroscopy~\cite{NuclearNoise,PhysRevB.101.235416}. In the steady state the correlator in the integrand does not depend on $t$. Its dependence on $\tau$ is given by the solution of Eq.~\eqref{eq:dIpm0}, which should be averaged over the initial conditions taken from the equilibrium nuclear spin distribution function.

\begin{figure}
  \includegraphics[width=1.0\linewidth]{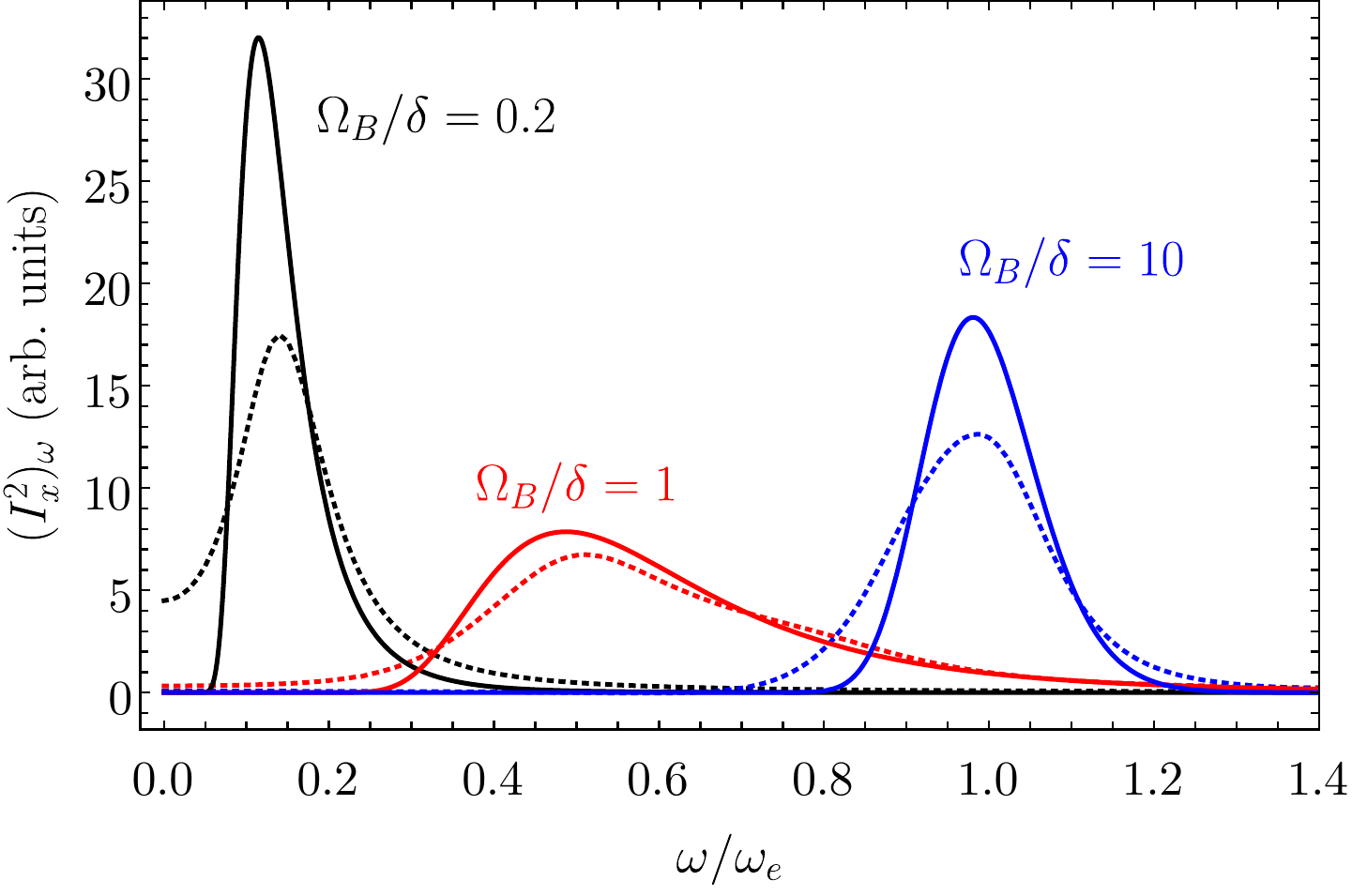}
  \caption{Nuclear spin noise spectra calculated after Eq.~\eqref{eq:spec_nucl_trans} for the different strengths of the magnetic field as indicated in the labels, neglecting the nuclear Zeeman splitting, $\omega_B=0$. The dashed curves are calculated for the same parameters with addition of the nuclear spin relaxation time $\tau_s^n\omega_e=25$~\cite{supp}.}
  \label{fig:Iz2}
\end{figure}

The noise spectrum of the transverse spin components reads~\cite{supp}
\begin{multline}
  \label{eq:spec_nucl_trans}
  (I_x^2)_\omega=\sum_{\pm}\frac{\sqrt{\pi}\delta^3}{16\omega_\pm^3\Omega_B}\exp\left[-\left(\frac{\Omega_B}{\delta}\right)^2\left(\frac{\omega_e^2}{\omega_\pm^2}+1\right)\right]
\\\times\left[\frac{2\omega_e\Omega_B^2}{\omega_\pm\delta^2}\cosh\left(\frac{2\omega_e\Omega_B^2}{\omega_\pm\delta^2}\right)-\sinh\left(\frac{2\omega_e\Omega_B^2}{\omega_\pm\delta^2}\right)\right]
\end{multline}
and $(I_y^2)_\omega=(I_x^2)_\omega$. Here $\delta$ is the typical fluctuation of $\Omega_N$ and $\omega_\pm=\omega\pm\omega_B$. This result agrees with the numerical calculations performed in Ref.~\onlinecite{PhysRevB.97.195311}. The nuclear spin noise spectrum is shown in Fig.~\ref{fig:Iz2} by the solid curves for the case of zero nuclear $g$-factor ($\omega_B=0$). Generally, the spectrum is an even function of $\omega$, so only the positive frequencies are shown in the figure. The spectrum consists of a single peak, which shifts from $\omega=0$ to $\omega=\omega_e$ with increase of the magnetic field. Its width changes nonmonotonously: it vanishes in the limits of weak and strong magnetic field, and it is of the order of $\omega_e$ when $\Omega_B\sim\delta$, the width of the peak is of the order of its central frequency in this case.

The shift of the peak in the spin noise spectrum with increase of the magnetic field is related to the acceleration of the nuclear spin precession in the Knight field. In small magnetic field, the electron spin is almost parallel to the nuclear spin, so it hardly causes the nuclear spin precession. However, the stronger the magnetic field, the larger the deviation of the average electron spin $\bar{\bm S}$ from the direction of the total nuclear spin $\bm I$, and the faster the nuclear spin precession. In the limit of strong magnetic field, the electron spin is parallel to it, which leads to the precession of the transverse nuclear spin components with the frequency $\omega_e$ (in the case of $\omega_B=0$). Hence, the nuclear spin noise spectrum is centered around this frequency~\cite{PhysRevB.97.195311}. The finite nuclear $g$ factor leads to the splitting of the peaks at both negative and positive frequencies~\cite{supp}.

The effect of the electron, $\tau_s^e$, and nuclear, $\tau_s^n$, spin relaxation times can be described using the kinetic equations for the distribution functions of $\bm I^\pm$~\cite{supp}. The effect of $\tau_s^n$ is illustrated in Fig.~\ref{fig:Iz2} by the dotted curves. The nuclear spin relaxation generally broadens the spectra. In particular, in weak ($\Omega_B\tau_s^n\omega_e/\delta\ll 1$) and strong ($\Omega_B/\delta\gg1$) magnetic fields the spectrum represents a Lorentzian at $\omega=0$ and $\omega=\omega_e$, respectrively, with the width $1/\tau_s^n$. Moreover, if the nuclear spin relaxation is fast, $\tau_s^n\omega_e\ll 1$, the spectrum is always Lorentzian centered at Zero frequency having the large width $1/\tau_s^n$.

\addDima{\section{Nuclear spin squeezing and entanglement}}As another important application, we describe the squeezing of the nuclear spin distribution function~\cite{MA201189}. The spin squeezing is widely studied nowadays~\cite{PhysRevLett.99.163002,Appel10960,PhysRevA.102.032612,Xiao2020} mainly in the field of quantum metrology as it allows one to increase the phase sensitivity in the Ramsey interferometry beyond the standard quantum limit~\cite{RevModPhys.90.035005}. In application to quantum dots it can be also used to increase the electron spin coherence time~\cite{PhysRevLett.107.206806,PhysRevLett.119.130503}. Previously, it was suggested that the nuclear spin squeezing can be produced by the quadrupole interaction~\cite{Bulutay2012} or in the presence of external driving under the condition of a fast electron spin relaxation~\cite{PhysRevLett.107.206806}.

\begin{figure}
  \includegraphics[width=\linewidth]{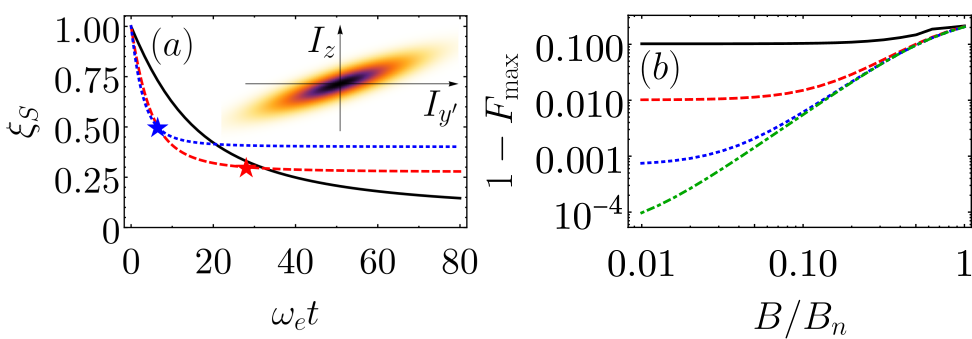}
  \caption{(a) Degree of the nuclear spin squeezing, $\xi_S$, as a function of free nuclear spin precession for the nuclear spin polarization $P=10\%$ (black solid curve), $30\%$ (red dashed curve) and $50\%$ (blue dotted curve). The inset shows the nuclear spin distribution in the direction perpendicular to the average spin at time marked by the blue star. External magnetic field equals to the average nuclear field, $\Omega_B=|\braket{\bm\Omega_N}|$ and $N=10^6$. (b) Infidelity of GHZ state preparation as a function of the applied magnetic field $\Omega_B/\Omega_N$ for $N=6$ (black solid curve), $80$ (red dashed curve), $1200$ (blue dotted curve), and $2\cdot10^4$ (green dot-dashed curve).}
  \label{fig:squeezing}
\end{figure}

Our solution of the nuclear spin dynamics predicts the dependence of the nuclear spin precession frequency on its value, Eq.~\eqref{eq:omega_n_f}. Therefore after the preparation of the coherent nuclear spin state with the average polarization being perpendicular to the external magnetic field (say, along the $x$ axis) and electron spin polarization along $\bm\Omega_e$~\cite{supp} different spins in the distribution precess with the different frequencies. This produces the nuclear spin squeezing \textit{intrinsically} in the central spin model. An example of the squeezed nuclear spin distribution produced in this way is given in the inset in Fig.~\ref{fig:squeezing}(a).

The distribution squeezing is described by the parameter $\xi_S$~\cite{PhysRevA.47.5138}, which is the ratio of the minimal spin standard deviation over the direction perpendicular to the average spin and its value for the coherent spin state~\cite{supp}. It is shown in Fig.~\ref{fig:squeezing}(a) as a function of time for the different nuclear spin polarization degrees, $P$. One can see that the larger the polarization is, the faster $\xi_S$ decreases, but the sooner it saturates. In typical GaAs based QDs, $\omega_e\sim1~\mu$s$^{-1}$ and spin relaxation time $\tau_s^n\sim0.1$~ms, so $\omega_e\tau_s^n\sim100$ and very strong nuclear spin squeezing, $\xi_S\sim10^{-2}$ can be reached.

The interferometry beyond the standard quantum limit requires the metrological degree of the spin squeezing $\xi_R=\xi_S/P$ to be less then unity~\cite{PhysRevA.46.R6797}. This criterion is more difficult to satisfy, and for $P=10\%$ it is not reached. However, for larger nuclear spin polarizations it is reached at the points marked by the red and blue stars in Fig.~\ref{fig:squeezing}(a). Since in modern experiments the polarization up to $80\%$ is feasible~\cite{Chekhovich_constants}, we believe that the metrological nuclear spin squeezing can be obtained as well.

Spin squeezing evidences the nuclear spin entanglement~\cite{RevModPhys.90.035005}. In the central spin model it is produced by the indirect interaction between nuclei mediated by the electron spin. While there is a number of entanglement measures~\cite{GUHNE20091}, an example of the maximally entangled state is the GHZ state, which is a coherent superposition of the collective spin states pointing in the opposite directions. In order to approach this state, we suggest to orient the electron spin in the direction, which is perpendicular to both initial nuclear spin direction direction and external magnetic field (say, the $y$ axis)~\cite{supp}. In this case the good electron spin quantization axis is perpendicular to it, so the nuclear spin dynamics represents the coherent superposition of precessions with the frequencies $\bm\omega_n^\pm$, see Eqs.~\eqref{eq:nucl_osc}. After the relative phase $(\omega_n^+-\omega_n^-)t$ reaches $\pi$, the nuclear spin state is close to GHZ state.

The infidelity \cite{RevModPhys.90.035005, GUHNE20091} of the GHZ state preparation is shown in Fig.~\ref{fig:squeezing}(b) as a function of the magnetic field for the different number of nuclei. It decreases with decrease of the magnetic field and increase of the number of nuclei, but generally it is very low. However, the smaller the magnetic field, the slower the increase of the relative phase. Since the preparation time should be below $\tau_s^n$, one has to consider $B/B_n\gtrsim10^{-2}$, which still produces very high fidelity up to $99.99\%$.

We note also that since the nuclear spin polarization produces macroscopic magnetic fields of the order of a few Tesla, the nuclear GHZ state would be a coherent superposition of the macroscopically different states, known as a Schrodinger cat state~\cite{schrodinger1935gegenwartige}. These state are important for quantum metrology and investigations of the quantum-classical correspondence~\cite{RevModPhys.85.1103,duan2019creating,Omran570}.

\section{Discussion and conclusion}The box model considered in this work is known to give qualitatively correct results also for the general central spin model~\cite{PhysRevLett.91.017402,PhysRevLett.104.143601,PhysRevB.99.035439,PhysRevB.102.085413}. For example, the nuclear spin noise spectra for homogeneous and inhomogeneous hyperfine coupling are very similar~\cite{PhysRevB.97.195311}. The most important deviations are expected in the nuclear spin squeezing and entanglement as the different nuclear spins would precess with the different frequencies for the inhomogeneous hyperfine coupling. However, the nuclear spin precession frequency would be almost the same in a central core of the electron wave function, which can include hundreds of nuclei. So the high degree of spin squeezing and entanglement is expected for these central spins.

The nuclear spin dynamics calculated in this work is important for the nuclear spin based quantum computations, as well as for the description of the optical properties of single quantum dots~\cite{Urbaszek,Chekhovich-review} and the transport properties of double quantum dots in the spin blockade regime~\cite{hanson07}. For example, the nuclear spin precession probably explains the low frequency peaks in the electron spin noise spectra predicted in the numerical simulations~\cite{noise-CPT}. Another application is related with organic semiconductors, where the hyperfine interaction determines the optical and electrical properties even at room temperature~\cite{KALINOWSKI2003710,bobbert,Nguyen2010}. We address this specific issue in a joint paper~\cite{noise_omar}.

In summary, in this work we derived the exact nonlinear equations for the nuclear spin dynamics and obtained their compact solution in the box model with many nuclear spins. It was used to calculate the nuclear spin noise spectra, and to describe the effects of nuclear spin squeezing and many body entanglement, which take place intrinsically after preparation of the appropriate coherent nuclear spin state. We believe that our results will be useful for the description of the electron and nuclear spin dynamics of the localized electrons in various nanostructures and different experimental conditions.


We gratefully acknowledge the fruitful discussions with M. M. Glazov and the partial financial by the RF President Grant No. MK-1576.2019.2 and Foundation for the Advancement of Theoretical Physics and Mathematics ``Basis''. We also thank S. G. Smirnov for design of Fig.~\ref{fig:precession_both}. The calculation of the nuclear spin dynamics by D.S.S. was supported by the Russian Science Foundation Grant No. 19-12-00051. A.V.S. acknowledges the support from the Russian Foundation for Basic Research Grant No. 19-02-00184.



\renewcommand{\i}{\ifr}
\let\oldaddcontentsline\addcontentsline
\renewcommand{\addcontentsline}[3]{}

\let\addcontentsline\oldaddcontentsline
\makeatletter
\renewcommand\tableofcontents{%
    \@starttoc{toc}%
}
\makeatother
\renewcommand{\i}{{\rm i}}

\onecolumngrid
\vspace{\columnsep}
\begin{center}
\newpage
\makeatletter
{\large\bf{Supplemental Material to\\``\@title''}}
\makeatother
\end{center}
\vspace{\columnsep}

The supplementary information presents the following topics:

\hypersetup{linktoc=page}
\tableofcontents
\vspace{\columnsep}
\twocolumngrid

\renewcommand{\section}[1]{\oldsec{#1}}
\renewcommand{\thepage}{S\arabic{page}}
\renewcommand{\theequation}{S\arabic{equation}}
\renewcommand{\thefigure}{S\arabic{figure}}
\renewcommand{\bibnumfmt}[1]{[S#1]}
\renewcommand{\citenumfont}[1]{S#1}

\setcounter{page}{1}
\setcounter{section}{0}
\setcounter{equation}{0}
\setcounter{figure}{0}


\section{S1. Comparison with numerical Hamiltonian diagonalization}
\label{sec:averages}

\begin{figure}[b]
  \centering
  \includegraphics[width=\linewidth]{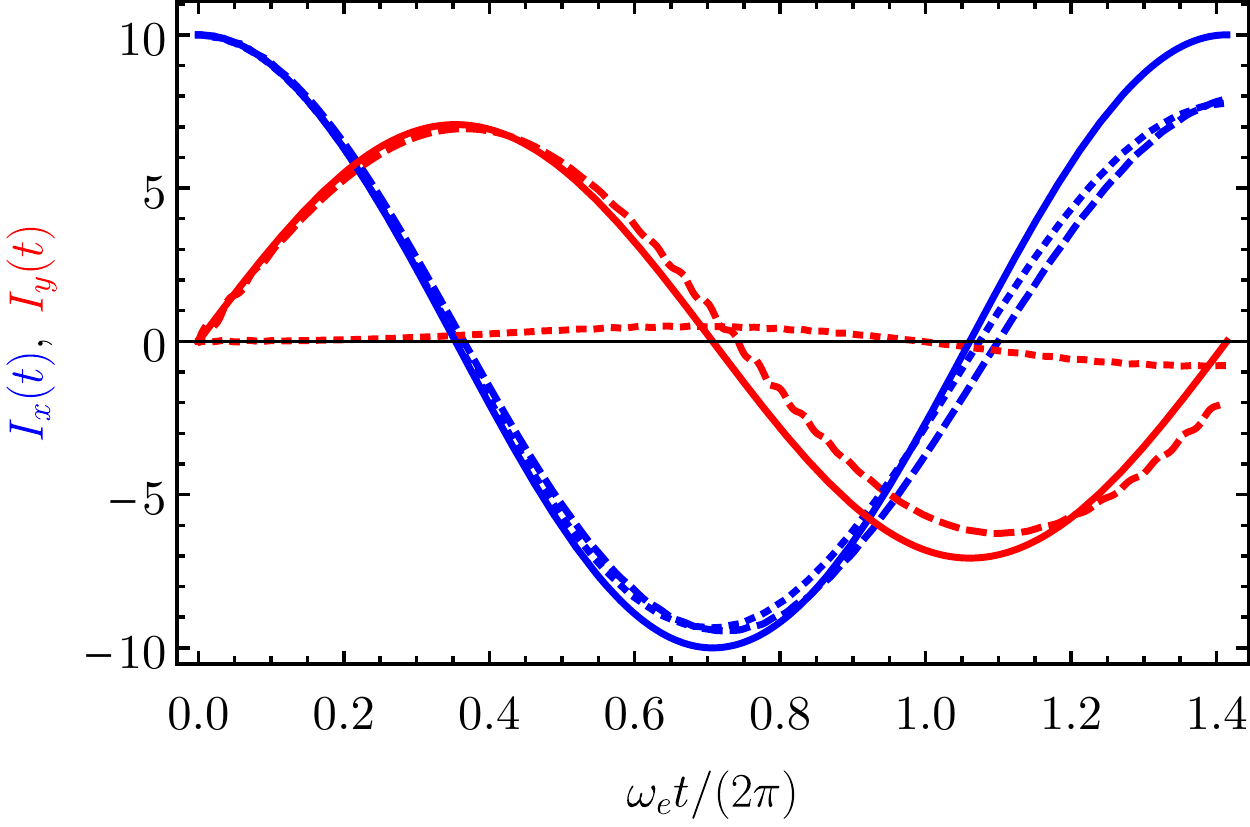}
  \caption{Dynamics of the nuclear spin components $I_x(t)$ (blue curves) and $I_y(t)$ (red curves) for the initial conditions $I_x(0)=I$, $I_y(0)=I_z(0)=0$. The solid curves are calculated after Eqs.~(12) of the main text for $\bm S(0)=\bm e_z/2$. The dashed curves are calculated numerically in the box model with $I=10$ for the same initial conditions. The dotted curves are the averaged numeric solutions for $\bm S(0)=\pm\bm e_z/2$, which corresponds to the unpolarized electron spin.}
  \label{fig:exact_prec}
\end{figure}

The comparison between our result and the exact solution of the Schrodinger equation for $I=10$ is shown in Fig.~\ref{fig:exact_prec}. Here the solid blue and red curves show $I_x(t)$ and $I_y(t)$ calculated after Eqs.~(12) of the main text for the initial conditions $\bm I(0)$ and $\bm S(0)$ parallel to the $x$ and $z$ axes, respectively, for $\Omega_B=A I/\hbar$. In this case $2(\bm\Omega_e\bm S)/\Omega_e=1/\sqrt{2}$, so the amplitude of the oscillations of $I_y(t)$ is $\sqrt{2}$ times smaller than that of $I_x(t)$ and $\omega_n=\omega_e/\sqrt{2}$. The solution of the Schrodinger equation is shown by the dashed curves and agrees reasonably well with the analytical expressions.

The dotted curves in Fig.~\ref{fig:exact_prec} show the averaged solution of the Schrodinger equation for the initial conditions $\bm S(0)=\pm\bm e_z/2$, which corresponds to the initially unpolarized electron spin. The dependence of $I_x(t)$ in this case is almost the same, while $I_y(t)$ is much smaller and has the opposite phase to $I_x(t)$. The quasiclassical Eqs.~(12) of the main text in this case yield the same $I_x(t)$ and $I_y(t)=0$ in agreement with the exact calculation.

Note that in the exact solution, the amplitude of the nuclear spin oscillations slowly decays with the rate ${\sim\omega_e/I}$. At the time scale $\sim I/\omega_e$ the nuclear spin recovers. This behaviour is not described by our model. In the limit $I\to\infty$ it disappears, so the exact solution and our result coincide.

Additionally, our theory is compared with the exact solution of the Schrodinger equation in Sec.~\hyperref[sec:GHZ]{S5}, where we show that Eqs.~(12) also correctly describe the coherent electron nuclear spin dynamics and predict formation of highly entangled states.


\section{S2. Role of the nuclear $g$ factor}
\label{sec:noise}

\begin{figure}[t]
  \centering
  \includegraphics[width=\linewidth]{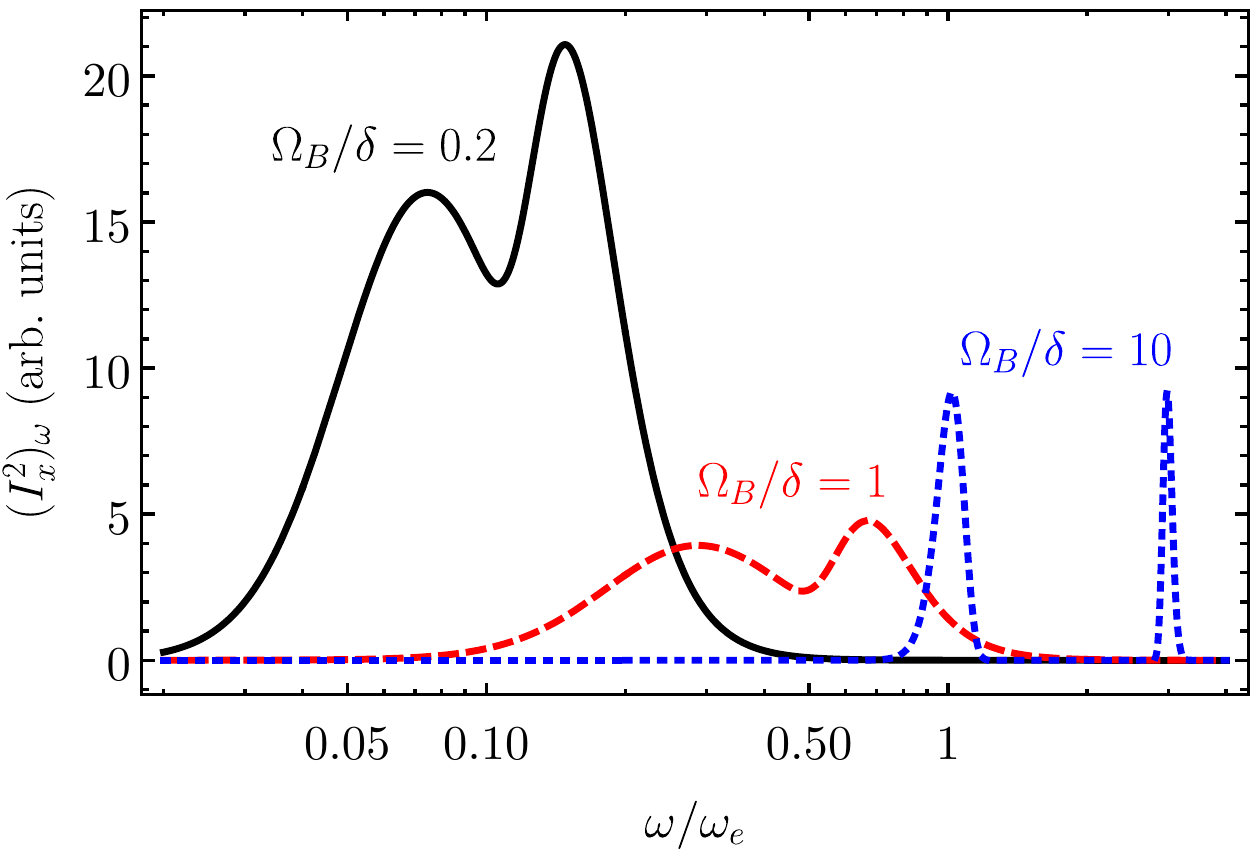}
  \caption{Nuclear spin noise spectra in the semi-logarithmic scale calculated after Eq.~(14) of the main text with the same parameters as for the solid lines in Fig.~(2) of the main text except for $\omega_B=0.2\Omega_B\omega_e/\delta$.}
  \label{fig:Iz2_gn}
\end{figure}

The role of the nuclear $g$ factor is illustrated in Fig.~\ref{fig:Iz2_gn} in the semi-logarithmic scale. As can be seen from Eq.~(14) of the main text, it leads to the splitting of the peaks at positive and negative frequencies by $2\omega_B$. As a result, the spectrum at positive frequencies consists of two peaks. In the strong field $\Omega_B\gg\delta$, the peaks are centered at the frequencies $\omega_B\pm\omega_e$. Qualitatively, this is caused by the nuclear spin precession in external magnetic field with the frequency $\omega_B$, which is increased or decreased by $\omega_e$ due to the electron spin polarization along or opposite to this direction~\cite{S_PhysRevB.97.195311} as follows from Eq.~(11) of the main text.

\section{S3. Role of the spin relaxation}
\label{sec:relaxation}

Our approach allows one to account phenomenologically for the electron and nuclear spin relaxations unrelated with the hyperfine interaction. Since the equations of the spin dynamics [Eqs.~(10) and~(11) of the main text] are nonlinear, it is necessary to introduce the probability distribution functions $f_\pm(t,\bm I)$ of $\bm I_\pm$, respectively~\cite{S_book_noise}. They are normalized by
\begin{equation}
  \int[f_+(t,\bm I)+f_-(t,\bm I)]\d\bm I=1,
\end{equation}
and satisfy the following phenomenological kinetic equations of the Fokker-Plank type:
\begin{equation}
  \label{eq:kinetic}
  \frac{\partial f_\pm}{\partial t}+\bm\nabla\left[\left(\bm\omega_n^\pm\times\bm I-\frac{\bm I}{\tau_s^n}\right)f_\pm\right]+D\Delta f_\pm+\frac{f_\pm-f_\mp}{\tau_s^e}=0,
\end{equation}
where $\bm\nabla=\partial/\partial\bm I$, $\Delta=\bm\nabla^2$, $\tau_s^{n,e}$ are the nuclear and electron spin relaxation times and $D=(\hbar\delta/A)^2/(2\tau_s^n)$ is an effective diffusion coefficient. We note that separately the nuclear spin relaxation alone can be described using the method of random Langevin forces, while the electron spin relaxation alone can be included phenomenologically in Eq.~(10) of the main text. However, both of them can be accounted for only using the spin distribution functions.

The steady state solution of Eqs.~\eqref{eq:kinetic} simply reads $f_\pm=f^{(0)}(\bm I)$, where
\begin{equation}
  f^{(0)}(\bm I)=\frac{1}{2(\sqrt{\pi}\delta_I)^3}\exp\left(-I^2/\delta_I^2\right),
\end{equation}
with $\delta_I=\hbar\delta/A$.

The spin noise spectrum is given by~\cite{S_ll10_eng}
\begin{equation}
  \label{eq:distr}
  (\delta I_\alpha^2)_\omega=2\Re\left[\sum_\pm\int S_\omega^\pm(\bm I) I_\alpha\d\bm I\right],
\end{equation}
where $S_\omega^\pm(\bm I)$ represent the solution of the following equations:
\begin{multline}
  -\i\omega S_\omega^\pm+\bm\nabla\left[\left(\bm\omega_n^\pm\times\bm I-\frac{\bm I}{\tau_s^n}\right)S_\omega^\pm\right]+D\Delta S_\omega^\pm+\frac{S_\omega^\pm-S_\omega^\mp}{\tau_s^e}\\
  =f^{(0)}(\bm I)I_\alpha.
\end{multline}
Below we analyze the role of the nuclear spin relaxation only. The role of the electron spin relaxation is studied in Ref.~\onlinecite{S_noise_OMAR}.

Since the nuclear spin precession in the Knight field does not change the total nuclear spin component along the $z$ axis, it monoexponentially decays on average with the rate $1/\tau_s^n$. As a result, the noise spectrum of $I_z$ has a simple Lorentzian form~\cite{S_PhysRevB.97.195311}
\begin{equation}
  (I_z^2)_\omega=\frac{\tau_s^n}{1+(\omega\tau_s^n)^2}\frac{\hbar}{A}
\end{equation}
with the width determined by the nuclear spin relaxation time. This spectrum is centered around zero frequency.

\begin{figure}[t]
  \centering
  \includegraphics[width=\linewidth]{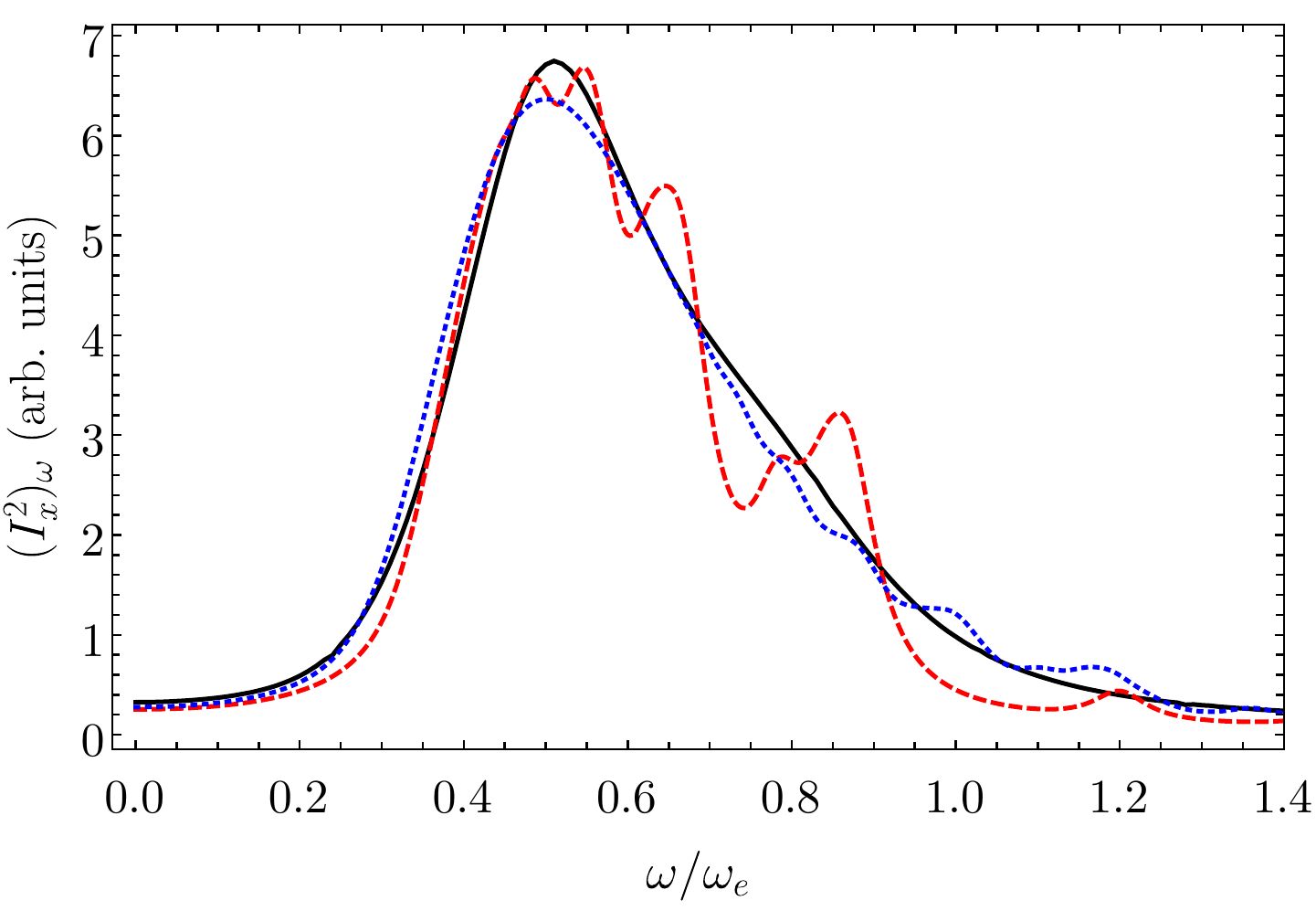}
  \caption{Nuclear spin noise spectrum calculated numerically for $\Omega_B=\delta$, $\omega_B=0$, $\tau_s^n\omega_e=25$, and $\tau_s^e=\infty$ (black solid curve). The red dashed and blue dotted curves are calculated following Ref.~\cite{S_PhysRevB.97.195311} for the same parameters and $I=10$ and $100$, respectively.}
  \label{fig:exact}
\end{figure}

It is instructive to compare our results with the previous calculation of the nuclear spin noise spectrum, which is based on the diagonalization of the Hamiltonian and summation of many contributions from the pairs of eigenstates~\cite{S_PhysRevB.97.195311}. The comparison is shown in Fig.~\ref{fig:exact} for the case of $\Omega_B=\delta$. In the limit of large $N$ our results coincide within the accuracy of our numerical computation. However for small $N$ the model of Ref.~\onlinecite{S_PhysRevB.97.195311} yields the oscillations in the spectrum, which are related to the finite number of the eigenstates of the system.

The model of Ref.~\cite{S_PhysRevB.97.195311} included the nuclear spin relaxation phenomenologically using the convolution with the Lorentzian with the width $1/\tau_s^n$. We checked that the difference with this model does not exceed 10\% even for the comparable spin relaxation time and spin precession period ($\Omega_B=\delta$, $\tau_s^n\omega_e=2.5$).

\section{S4. Nuclear spin squeezing}

The degree of the nuclear spin distribution squeezing for $N$ spin $1/2$ particles is defined as follows~\cite{S_PhysRevA.47.5138}:
\begin{equation}
  \label{eq:xi_s}
  \xi_S=\frac{2\min_{\bm n}\sqrt{\braket{I_n^2}}}{\sqrt{N}},
\end{equation}
where the minimum is taken over the directions $\bm n$ perpendicular to the average spin $\braket{\bm I}$ and $I_n=\bm n\bm I$. If all the nuclei are independent and have the average spin along $\braket{\bm I}$, then $\xi_S=1$. Thus $\xi_S<1$ indicates the squeezing of the nuclear spin distribution with respect to its typical symmetric state.

Our solution of the central spin box model suggest the following possible way of the squeezed spin state generation: (i) Nuclei can be dynamically polarized along the magnetic field (the $z$ axis). (ii) Radiofrequency $\pi/2$ pulse can be applied to bring the average nuclear spin into $(xy)$ plane. (iii) The electron spin can relax to its eigenstate parallel to $\bm\Omega_e$. (iv) Then the slow nuclear spin precession in the Knight field will produce squeezed nuclear spin state.

To be specific, in the main text we consider the initial nuclear spin distribution function
\begin{equation}
  f_+(\bm I,0)=\frac{1}{\epsilon(\sqrt{\pi}\delta_I)^3}\exp\left[-\frac{(I_x-PN/2)^2}{(\epsilon\delta_I)^2}-\frac{I_y^2+I_z^2}{\delta_I^2}\right],
\end{equation}
where $\epsilon=\sqrt{1-P^2}$ describes the suppression of the nuclear spin fluctuations along the polarization direction with $P$ being the nuclear spin polarization degree~\cite{S_PolarizedNuclei}; and $f_-(\bm I,0)=0$. We note that $\delta_I^2=N/2$.

Then the solution of Eq.~\eqref{eq:kinetic} yields the time evolution of the nuclear spin distribution function. For simplicity we neglect the electron and nuclear spin relaxations as well as the nuclear Zeeman splitting ($\omega_B=0$). In this limit the kinetic equation can be simply solved analytically with the result $f_-(\bm I,t)=0$ and
\begin{equation}
  f_+(\bm I,t)=f_+(\mathcal R_t^{-1}(\bm I),0).
\end{equation}
Here the operator $\mathcal R_t(\bm I)$ describes the spin rotation about the $z$ axis by an angle $\omega_n^+t$. Due to the dependence of the precession frequency $\omega_n^+$ on $\bm I$, the nuclear spin distribution is squeezed for $t>0$.

To illustrate this we consider a given time $t$ and denote the average spin direction as $x'$. Note that this axis lies in the $(xy)$ plane. The nuclear spin distribution function in the $(y'z)$ plane is given by
\begin{equation}
  \tilde f(I_y',I_z,t)=\int f_+(I_{x'},I_{y'},I_z,t)\d I_{x'}.
\end{equation}
This distribution function is shown in the inset in Fig.~3(a) in the main text. Its deviation from the disk shows the nuclear spin squeezing.

The degree of the nuclear spin squeezing can be calculated directly from the definition~\eqref{eq:xi_s} using the distribution function $f_+(\bm I,t)$. It is shown in Fig.~3 of the main text for the parameters given in the caption. It is noteworthy that the absolute value of the nuclear spin polarization does not change significantly at this time scale.


\section{S5. Nuclear spin entanglement}
\label{sec:GHZ}

For an ensemble of independent nuclear spins one has $\xi_S=1$. Thus $\xi_S<1$ evidences at least the pairwise entanglement between the nuclear spins. This entanglement is produced by the indirect interaction between nuclei mediated by the electron. Ones of the most entangled nuclear spin states are the Greenberger Horne Zeilinger (GHZ) states, which have $N$ particle entanglement. A family of these state can be defined by the following wave functions:
\begin{equation}
  \Psi_{GHZ}(\varphi)=\dfrac{\prod\limits_{k=1}^N\left(^{1/\sqrt{2}}_{i/\sqrt{2}}\right)_k+\e^{\i\varphi}\prod\limits_{k=1}^N\left(^{1/\sqrt{2}}_{-i/\sqrt{2}}\right)_k}{\sqrt{2}},
\end{equation}
where $\varphi$ is an arbitrary phase, the spinors $\left(^{1/\sqrt{2}}_{\pm i/\sqrt{2}}\right)_k$ describe the state of the $k$-th nuclei with the spin along or opposite to the $y$ axis, and the symbol $\prod$ denotes the direct product. Thus the states $\Psi_{GHZ}$ represent the coherent superposition of the states with the total nuclear spin $I=N/2$ being directed along and opposite to the $y$ axis.

It follows from Eqs.~(12) that the GHZ states can be produces from the initial coherent nuclear spin state
\begin{equation}
  \Psi(0)=\prod\limits_{k=1}^N\left(^{1/\sqrt{2}}_{1/\sqrt{2}}\right)_k,
\end{equation}
which describes the complete nuclear spin polarization along the $x$ axis. This state may be prepared following the protocol described in the previous section. However, to generate the GHZ state, the electron spin should be initially oriented perpendicular to $\bm\Omega_e$, for example, along the $y$ axis. In this case the probabilities to find it parallel and antiparallel to the good quantization axis $\bm\Omega_e$ are $P_+=P_-=1/2$. As a result, the nuclear spin $\bm I$ precesses with the opposite frequencies $\omega_n^\pm$ about the $z$ axis. As a result after the time $t_\pi=\pi/(2|\omega_n^\pm|)$ the total nuclear spin is approximately in the GHZ state.

We calculated the fidelity of the GHZ state preparation by exact solution of the Schrodinger equation with the Hamiltonain (1) in the main text. The fidelity was calculated as
\begin{multline}
  F_{\rm max}=\max_\varphi\left(\left|\braket{\Psi_{GHZ}(\varphi),\uparrow|\Psi(t_\pi)}\right|^2\right.\\\left.+\left|\braket{\Psi_{GHZ}(\varphi),\downarrow|\Psi(t_\pi)}\right|^2\right),
\end{multline}
where $\Psi(t_\pi)$ denotes the total system wave function at time $t_\pi$ calculated with the initial conditions described above, and $\ket{\Psi_{GHZ}(\varphi),\uparrow/\downarrow}$ denote the wave functions with the nuclear spin state $\Psi_{GHZ}(\varphi)$ and the electron spin parallel/antiparallel to the $z$ axis. The fidelity shown in Fig.~3(b) of the main text reaches $99.99\%$, which additionally proves the applicability of our results to the description of the dynamics of quantum coherent electron nuclear spin states.

Finally, we note that the $\pi/2$ pulses, which tilt the average nuclear spin from the $z$ axis to the $(xy)$ plane have the length, which can be of the order of microsecond. In order to make them more efficient, the electron can be removed for this time from the quantum dot, which is easy in the gated structures. Moreover, in the above description we neglected the nuclear Zeeman splitting, but if it is present, it would lead to the total nuclear spin precession about the $z$ axis with a constant frequency. So this would not affect neither nuclear spin squeezing not generation of the GHZ states.

\end{document}